\documentclass[12pt]{article}
\usepackage{amssymb}
\usepackage{amsmath}
\usepackage{graphicx}
\usepackage{cite}
\usepackage{float}
\usepackage{authblk}

\begin{document}


\title{Theoretical study of strain-dependent optical absorption in Stranski-Krastanov grown InAs/InGaAs/GaAs/AlGaAs quantum dots}

\author[1*]{\underline{Tarek Ameen}}
\author[1*]{\underline{Hesameddin Ilatikhameneh}}
\author[1]{Yuling Hsueh}
\author[1]{James Charles}
\author[1]{Jim Fonseca}
\author[1]{Michael Povolotskyi}
\author[2$\dagger$]{Jun Oh Kim}
\author[2]{Sanjay Krishna}
\author[3]{Monica S. Allen}
\author[3]{Jeffery W. Allen}
\author[4]{Brett R. Wenner}
\author[1]{Rajib Rahman}
\author[1]{Gerhard Klimeck}
\affil[1]{\normalsize{Network for Computational Nanotechnology, Department of Electrical and Computer Engineering, Purdue University, West Lafayette, IN 47907, USA}}
\affil[2]{\normalsize{Center for High Technology Materials, University of New Mexico, Albuquerque NM 87106.}}
\affil[3]{\normalsize{Air Force Research Laboratory, Munitions Directorate, Eglin AFB, FL 32542.}}
\affil[4]{\normalsize{Air Force Research Laboratory, Sensors Directorate, Wright-Patterson AFB, OH 45385.}}
\affil[$\dagger$]{\normalsize{now at Korean Research Institute of Standards and Sciences.}}
\affil[*]{\normalsize{These authors contributed equally to this work.}}
\renewcommand\Authands{ and }
\date{}
\maketitle
\providecommand{\keywords}[1]{\textbf{\textit{Keywords---}} #1}

\begin{abstract}
A detailed theoretical study of the optical absorption in self-assembled quantum dots is presented in this paper. A rigorous atomistic strain model as well as a sophisticated electronic band structure model are used to ensure accurate prediction
of the optical transitions in these devices .  The optimized models presented in this paper are able to reproduce the experimental results with an error less than 1$\%$.  The effects of incident light polarization, alloy mole fraction, quantum dot dimensions, and doping have been investigated. The in-plane polarized light absorption is more significant than the perpendicularly polarized light absorption. Increasing the mole fraction of the strain controlling layer leads to a lower energy gap and larger absorption wavelength. Surprisingly, the absorption wavelength is highly sensitive to changes in the dot diameter, but almost insensitive to changes in the dot height. This unpredicted behavior is explained by sensitivity analysis of different factors which affect the optical transition energy. 
\end{abstract}


\section{Introduction}

Self-assembled quantum dots have improved the performance of many optoelectronic devices, such as quantum dot infrared photodetectors (QDIPs)\cite{liu79,tqd}, and intermediate band solar cells (IBSCs)\cite{nextgen,luque2012understanding}. QDIPs  have lower dark current than conventional photodetectors \cite{tqd} and are sensitive to normally incident light unlike their counterparts that are made from quantum wells \cite{tqd}. For IBSCs, one of the most successful methods in pushing solar cell efficiency beyond the Shockley-Queisser limit is to add one or more intermediate bands inside the gap, which can be realized by using quantum dots and quantum dot-in-a-well devices\cite{luque2012understanding}. 

The absorption coefficient $\alpha(\lambda)$ of quantum dots is an important parameter that needs to be well designed for the proper operation of these devices, and an accurate  model for the absorption coefficient $\alpha(\lambda)$  would be highly appreciated by researchers as well as engineers working in these fields. For this reason,  this study has been devoted to fill in some of the gaps found in the available models of the absorption, specially in the atomistic strain calculations that are needed for accurate description of the bound states. In addition, a detailed study is provided of the effects of varying doping, dimensions, alloy mole fraction, and incident light polarization.

Self-assembled quantum dots are highly strained heterostructures, the atomistic strain in such structures is usually on the order of 10$\%$. Such high strain values are beyond the domain of validity of the continuum elasticity theory \cite{pryor1998comparison}. Effects of interface inter-diffusion and alloy disorder further complicate the physics of determining the relaxed atom positions and a rigorous atomistic treatment of strain is needed (such as Keating \cite{keating1966} or anharmonic models\cite{laz_2003}). 

The anharmonic strain model \cite{laz_2003} reported by Lazarenkova, et. al. modified the standard harmonic Keating model to include the effect of anharmonicity in the lattice potential. This modification is necessary as the harmonic Keating potential underestimates the repulsive forces at smaller atomic separations and fails to reproduce the weakening of the strain energy at large bond lengths \cite{laz_2004}. The anharmonic strain model of Lazarenkova et. al. \cite{laz_2003} has five strain parameters for optimization compared to just two parameters in the harmonic Keating model.
The anharmonic strain parameters were originally optimized to obtain correct Gr{\"u}neisen parameters for more accurate phonon dispersion calculations \cite{laz_2003}, however, it will be shown later that the original parameter set cannot reproduce the experimental optical absorption peaks in quantum dots \cite{tate}.
Moreover, these parameters do not capture the correct strain distribution in quantum wells, for which an analytical solution is known. The approach described in this paper for optimizing the parameters to obtain correct biaxial strain ratio in the quantum wells has resulted in an improvement in quantum dot simulations in terms of capturing experimental optical transitions much more accurately than previously. 
The Hamiltonian has been constructed with semi-empirical tight binding with 20-orbital sp3d5s* basis per atom,
including spin-orbit interaction (sp3d5s*\textunderscore SO) \cite{boykin2002diagonal}. The absorption coefficient has been calculated by employing Fermi's golden rule.  
 
In section II, the systems simulated and the numerical tools used are described. In section III, the theoretical aspects of the problem and the optimization of the strain model is discussed.  Lastly in section IV, the results of the simulation is discussed in addition to analyzing the sensitivity of the absorption to various quantum dot parameters.

\section{Multi Million Atom Simulation}
As shown in the Figure \ref{SKQD_fig}, the investigated SK-QD system \cite{barve2012barrier} has a dome shaped InAs quantum dot with a base diameter of 20 nm and a height of 5 nm. The wetting layer is 2 monolayers.  The measured system has been doped with sheet doping of two electrons per dot. The strain controlling layer is made of In$_{0.15}$Ga$_{0.85}$As and is sandwiched between two layers of GaAs each with a thickness 1 nm. Next, there are two layers of  Al$_{0.22}$Ga$_{0.78}$As, each with a thickness of 2 nm. The rest of the structure is made of Al$_{0.07}$Ga$_{0.93}$As. The dimensions of the simulated SK-QD systems are 60 nm x 60 nm x 60 nm. The strain simulation contains around 10 Million atoms and the atomistic grid is shown in Figure \ref{Atomistic_fig}. The band structure calculations do not need all of these atoms, since bound states decay exponentially outside the quantum dot. The band structure calculations are performed on 40 nm x 40 nm x 20 nm  box surrounding the quantum dot, this box contains only 1.5 million atoms.  Each atom has 20 orbitals in the sp3d5s*\textunderscore SO tight binding basis.  Such large systems are computationally expensive to run strain and electronic structure simulations and require highly scalable computational codes, the described system would take around three hours on 729 cores. The code that has been used for our simulations is the Nano Electronic MOdeling tool version 5 ''NEMO5'' \cite{nemo5}.

\section{Theoretical Model}
\subsection{Atomistic strain model}
The Harmonic Keating  strain model \cite{keating1966} treats the interatomic forces as spring forces connecting the atoms together. It considers nearest neighbor interactions only and the expression for the elastic energy is given by
\begin{equation}
E=\frac{3}{8}\sum\limits_{m,n}{\left[ \frac{{{\alpha }_{mn}}}{d_{mn}^{2}}{{\left( r_{mn}^{2}-d_{mn}^{2} \right)}^{2}}+\sum\limits_{k>n}{\frac{{{{\beta }_{mnk}}}}{{{d}_{mn}}{{d}_{mk}}}\left( {{r}_{mn}}\cdot {{r}_{mk}}-{{d}_{mn}}\cdot {{d}_{mk}} \right)} \right]} , 
\label{eqn_1}
\end{equation}
where the coefficient $\alpha$ corresponds to the force constant of the bond length distortion, and $\beta$ corresponds to the bond angle distortion as shown in Figure \ref{interactions_fig}. $\alpha$ and $\beta$ are material-dependent constants. $r_{mn}$ is the displacement vector from atom m to atom n for the strained crystal, while $d_{mn}$  is the same vector for the bulk unstrained crystal. The summation is taken over the nearest neighbors only and the total energy is minimized with respect to the individual atomic positions, relaxing the structure. The problem with the harmonic Keating potential given by equation (\ref{eqn_1}) is that, it produces a symmetric energy profile around the equilibrium interatomic distance and angle. For this reason, the Keating model fails to reproduce the weakening of the strain energy with increasing bond length and it underestimates the repulsive forces at close atomic separation \cite{laz_2003,laz_2004}.
The anharmonic correction of the Keating model proposed by Lazarenkova, et. al. \cite{laz_2003,laz_2004} has been able to solve this problem by modifying the two parameters $\alpha$ and $\beta$ of the Keating model and making them functions of bond length $r$ and bond angle $\theta$ as given by
\begin{multline}
{{\alpha }_{mn}}=\alpha _{mn}^{0}\left( 1-{{A}_{mn}}\frac{r_{mn}^{2}-d_{mn}^{2}}{d_{mn}^{2}} \right), 
\\
{{\beta }_{mnk}}=\beta _{mnk}^{0}\left( 1-{{B}_{mnk}}\left( \cos \left( {{\theta }_{mnk}} \right)-\cos \left( \theta _{mnk}^{0} \right) \right) \right)\times \left( 1-{{C}_{mnk}}\frac{{{r}_{mn}}\cdot {{r}_{mk}}-{{d}_{mn}}\cdot {{d}_{mk}}}{{{d}_{mn}}\cdot {{d}_{mk}}} \right). 
\label{eqn_2}
\end{multline}
In this anharmonic strain model, there are five strain parameters for each material, $\alpha _{mn}^{0}$, $\beta _{mnk}^{0}$, ${A}_{mn}$, ${B}_{mnk}$, and ${C}_{mnk}$. The anharmonic model was developed mainly for simulating phonon dispersion and transport. For this reason, the anharmonic strain parameters were optimized to reproduce the Gr{\"u}neisen parameters $\gamma_i=-\frac{V}{\omega_i} \frac{\delta \omega_i}{\delta V}$ which are a measure of the dependence of the phonon mode frequencies on strain. Simulating the strain in quantum dots with the original anharmonic strain parameters provide inaccurate results that do not match experimental results as shown in the results section of the paper. 

In addition, simulating strain in quantum wells with these parameters gives strain tensor components that do not match well the analytical solution of the strain in quantum wells as shown in Table \ref{Table}. The parameters of the model described in this paper have been tuned to reproduce the biaxial strain ratio $\sigma$  of InAs in order to capture the strain distribution in quantum wells and quantum dots.  The InAs biaxial strain ratio $\sigma $ of 1.053 was obtained from more accurate but size-limited density functional theory calculations \cite{hammerschmidt2007elastic}.
Only one parameter needs to be optimized in this approach. Out of the five strain parameters, tuning the parameter $\alpha^0$ while keeping the other parameters as reported in \cite{laz_2003} yields the best convergence and stability of the strain model. The new value of $\alpha^0$ after optimization is 19.35Nm$^{-1}$, which can be readily used in accurate large-scale atomistic strain simulations. Table \ref{Table} shows the atomistic strain calculated for InAs/GaAs quantum well, as shown in Figure \ref{QW_fig}, before and after tuning. The analytical expressions for the strain components in quantum wells are $\epsilon_{||}=\frac{a_{GaAs}-a_{InAs}}{a_{InAs}}$ and $\epsilon_{\bot}= - \sigma \epsilon_{||}$ \cite{sun2010strain}, where $a$ is the lattice constant.

\begin{table}
\centering
\begin{tabular} { |c|c|c| }
  \hline
  \textbf{Method} & $\mathbf{\epsilon_{||}}$ & $\mathbf{\epsilon_{\bot}}$\\ \hline
  Analytical                & - 6.68 $\%$  & 7.04 $\%$ \\ \hline
  Anharmonic before tuning  & - 6.68 $\%$  & 8.9 $\%$ \\ \hline
  Anharmonic after tuning   & - 6.68 $\%$  & 7.04 $\%$ \\
  \hline
\end{tabular}
\caption{Strain calculated for InAs/GaAs quantum well. Tuning has improved the anharmonic strain results in quantum well.}
\label{Table}
\end{table}

\subsection{Electronic structure and absorption}
The eigenstates of the system were calculated with a Hamiltonian constructed with semi-empirical tight binding sp3d5s*\textunderscore SO basis. The Slater-Koster TB \cite{slater1954simplified} parameters for InAs, GaAs, and AlAs are taken from \cite{klimeck2002development,boykin2007approximate}. For including the effect of strain on the tight binding Hamiltonian, please refer to \cite{boykin2002diagonal}. These parameters are well established and have been verified before versus experimental measurements of quantum dots \cite{klimeck2007atomistic,usman,neupane2015effect,neupane2011core}.

For the absorption coefficient $\alpha$, Fermi golden rule has been used to calculate the absorption coefficient\cite{Tarek,Tarek2}, 
\begin{equation}
\alpha(\omega)=\frac{2~ \pi~ \omega ~ n_{dots}}{\acute{n}~ \epsilon_0 ~ c  } \sum\limits_{i,f} \left |\underline{d_{fi}} \cdot \underline{\hat{e}} \right|^2 \delta(E_f-E_i-\hbar \omega) (F_i-F_f) ~ ,
\label{eqn_3}
\end{equation}

where $n_{dots}$ is the number of quantum dots per unit volume, $\omega$ is the photon angular frequency, $E_i$ and $E_f$ are initial and final energies of the transition, $F_i$ and $F_f$ are occupation probability of the initial and final states, $\acute{n}$ is the refractive index of the material, $c$ is the speed of light in free space, $\epsilon_0$ is the free space permittivity, $\underline{\hat{e}}$ is the polarization of the incident light, and $\underline{d_{fi}}$ is the first order dipole moment that is given by $\underline{d_{fi}}=q <\psi_f|\underline{r}|\psi_i> $, where $q$ is the electron charge.

For the transitions between the valence bound states and the conduction bound states, $F_i=1$, where $F_f$ depends on the energy level and the doping. Normally, quantum dots are occupied by the number of electrons equal to the average number of dopants per dot\cite{ameen2014modeling}. This approach is reasonable for quantum dots that are far from heavily doped regions, however, it is not appropriate for quantum dots adjacent to heavily doped regions, such as contacts. In addition, it is assumed that if one level is filled then the next electron fills a higher level instead of filling the same level with opposite spin due to the high Coulomb repulsion between them. Irrespective of the filling, the Coulomb repulsion shift in the energy has been neglected. In other words if the doping is two electrons per dot, then the electron ground state and the first excited state will be occupied each by an electron and they will not accept more electrons. These assumptions have been considered while calculating the absorption.

\section{Results and Discussion}
\subsection{Simulation versus Experimental Results}
A comparison with the measured absorption spectrum \cite{barve2012barrier} of the SK-QD system has been made to validate the model. Figure \ref{Experimental_fig} shows the calculated and measured absorption spectrum of the device. The simulation result matches very well with the measured absorption and the error in estimating the energy of the absorption peak is less than 1 $\%$. Further comparison with experimental measurements are provided later when discussing the effect of the alloy mole fraction of the strain controlling layer.

\subsection{Band structure and states}
Figure \ref{States2_fig} shows the wavefunction probability density of the first eight non-degenerate states of both electrons and holes. It is worth noting that the hole ground state has an s orbital like shape.

SK-QDs have complicated band profile as multiple effects, such as geometric confinement, strain, alloy disorder, etc. can cause major changes in the band edges of the bulk material. It is important to know where the wavefunctions of the electrons and holes are localized due to these disordered band edges, as the spatial overlap between the states will determine the optical absorption spectrum. Hence, one can look at the conduction and valence band edges along arbitrary lines passing through the quantum dot. This can be done by using the deformation potential theory that gives the shift of the band edges due to small lattice deformations. The shift in the band edges due to lattice strain for zincblende materials is given by reference \cite{bir1974symmetry}
\[\Delta {{E}_{c}}={{a}_{c}}{{\varepsilon }_{H}}_{{}},\]
\begin{equation}
\Delta {{E}_{vHH}}={{a}_{v}}{{\varepsilon }_{H}}+\frac{b}{2}{{\varepsilon }_{B}}_{{}},
\label{eqn_4}
\end{equation} 
\[\Delta {{E}_{vLH}}={{a}_{v}}{{\varepsilon }_{H}}-\frac{b}{2}{{\varepsilon }_{B}}_{{}}, \]
where $\Delta {{E}_{c}}$ is the shift in the conduction band edge, $\Delta {{E}_{vHH}}$ and  $\Delta {{E}_{vLH}}$ are the shifts in the heavy and light hole band edges, respectively. $a_c$, $a_v$, and $b$ are the deformation potential coefficients of the material. In these simulations, we have used the parameters recommended for III-V materials by \cite{vurgaftman5815}. ${\varepsilon }_{H}$ and ${\varepsilon }_{B}$ are the hydrostatic and biaxial strain components which are linear combinations of the atomistic strain components:  ${\varepsilon }_{H}= {\varepsilon }_{xx}+{\varepsilon }_{yy}+{\varepsilon }_{zz}$ and ${\varepsilon }_{B} = {\varepsilon }_{xx}+{\varepsilon }_{yy}-2{\varepsilon }_{zz}$ \cite{bir1974symmetry}, where z is the growth direction. Figure \ref{LBS12_fig} shows the band edges along two lines through the middle of the quantum dot along the [001] and [110] directions. The unstrained band edges are plotted to show the significant effect of strain on the band edges.

\subsection{Doping and Polarization}
Figure \ref{InPlanePerpendicular_fig} shows the absorption spectrum of the SK-QD with different cases of doping for in-plane polarized light (normally incident light) and perpendicularly polarized light. The in-plane polarized light absorption is more than ten times larger than the perpendicularly polarized light absorption. Quantum dots can absorb both polarization components, but their main advantage over quantum wells are the dots' sensitivity to normally incident light which makes them more efficient than quantum wells in applications like solar cells and photodetectors. For this reason, we will focus more on studying the in-plane polarized absorption properties. In Figure \ref{InPlanePerpendicular_fig}, each peak is coming from transitions between valence states and each conduction state. 
Practically, these peaks can overlap with each others if the inhomogeneous broadening, due to variance in the dot dimensions, is high. 

\subsection{Effect of Quantum Dot Dimensions}
Figure \ref{DiameterHeight_fig} shows the effect of changing quantum dot diameter and height on the in-plane polarized absorption spectrum. Increasing the dot diameter results in red-shifting the peaks, while increasing the dot height doesn't have a significant effect on wavelength.
In contrast, to the simple particle in a box problem, which predicts a stronger sensitivity to the smaller dimension (the height), this work shows that absorption wavelength is much more sensitive to changing the dot diameter than the dot height.

The effects of changing the dimensions on the energy transition $\Delta E$ between the hole and the electron ground states can be understood with a simple analytical model. This transition has two contributions: strain and confinement. The strain shifts the band edges and affects the energy gap $E_g$ , while the confinement increases the minimum allowed energy of electron $E_{elec}$ and hole $E_{hole}$ with respect to band edges. Let $E_{box} = E_{elec} + E_{hole}$, then the transition energy $E$ is
\begin{equation}
\Delta E  = E_g+ E_{box}.
\label{eqn_7}
\end{equation}
Due to the sign of the deformation potential and strain, the valence band edge inside the quantum dot is the heavy hole , from equation (\ref{eqn_4})
\begin{equation}
 E_g = E_{g,bulk} + (a_c-a_v) \varepsilon_H - \frac{b}{2} \varepsilon_B .
\label{eqn_8}
\end{equation}

Figure \ref{HB_DH_fig} shows the effect of changing the diameter and the height on the hydrostatic and biaxial strain. The magnitude of the biaxial strain increases with increasing the diameter and decreases with increasing the height, while the magnitude of the hydrostatic strain changes slightly in the opposite direction to the biaxial strain.
Increasing the height is equivalent to decreasing the diameter in terms of changing the strain in the quantum dot. Increasing the diameter reduces the energy gap which further reduces the optical transition energy, while increasing the height increases the energy gap which works against the reduction in the confinement energy. This compensation results in almost the same optical transition energy.  

Although the variations in the hydrostatic strain are smaller than the variations in the biaxial strain, as shown in Figure \ref{HB_DH_fig}, the hydrostatic strain variations shouldn't be neglected. This is due to the higher deformation potential weight for the hydrostatic strain. For example, $a_c-a_v = - 6 eV$ is six times higher than $\frac{b}{2} = -1 eV$ for InAs. Also, changing one of the dimensions either increases or decreases the hydrostatic strain, and it will have the opposite effect on the biaxial strain (decreases or increases), but the hydrostatic and biaxial strain  will work together in the same direction on the energy gap since their terms have opposite sign in equation (\ref{eqn_8}).

To get an expression for the $E_{box}$, the dome shaped quantum dot is approximated to be a disc of cylinder $R$ and height $H$, one can easily obtain $E_{box}$ by solving an effective mass Hamiltonian in the cylindrical coordinates,
\begin{equation}
 E_{box} = \frac{\hbar^2}{2} \left(\frac{1}{m_e}+\frac{1}{m_h}\right) \left(\frac{\pi^2}{H^2}+\frac{X_{01}^{2}}{m_h}\right)  ,
\label{eqn_9}
\end{equation}
where $m_e$, $m_h$ are the electron and the heavy hole effective masses, and $X_{01}=2.405$ is the first zero of Bessel function of the first kind with order 0. It should be noted that effective masses for the electron and heavy hole under strain are different from the bulk, and  the InAs effective masses are $m_e=0.1m_0$ and $m_h=0.48m_0$ \cite{klimeck2002development}.

%
%
%
%
For a quantum dot of $D=20$ nm and $H=5$ nm, the sensitivity of the confinement energy to the dot radius is $\frac{\delta E_{box}}{\delta R} =   - \hbar^2 X_{01}^2 \left(\frac{1}{m_e}+\frac{1}{m_h}\right) \frac{1}{R^3} \approx -5$ meV/nm, and the sensitivity of the energy gap to the dot radius is  $\frac{\delta E_g}{\delta R} = (a_c-a_v) \frac{\delta \varepsilon_H}{\delta R} - \frac{b}{2} \frac{\delta \varepsilon_B}{\delta R} \approx -17$ meV/nm  which give a total sensitivity of the optical transition $\frac{\delta \Delta E}{\delta R} \approx -22 $ meV/nm. 
 As for changing the height, in the same way $\frac{\delta E_{box}}{\delta H} =  - \hbar^2 \pi^2 \left(\frac{1}{m_e}+\frac{1}{m_h}\right) \frac{1}{H^3} \approx -68$ meV/nm, $\frac{\delta E_g}{\delta H}  = (a_c-a_v) \frac{\delta \varepsilon_H}{\delta H} - \frac{b}{2} \frac{\delta \varepsilon_B}{\delta H} \approx 65$ meV/nm  which give $\frac{\delta \Delta E}{\delta H} \approx -3$ meV/nm. For increasing the dot radius, both contributions reduce the transition energy. While increasing the dot height, both contributions are working against each other which reduces the sensitivity of the transition energy to the dot height. 
 
\subsection{Strain Controlling Layer}
Changing the In mole fraction of the InGaAs strain controlling layer (capping layer) is a convenient way to tune the absorption peak. The effect of mole fraction has been studied on a  slightly different system, reported in ref. \cite{tate}, which help us further validate the results of the simulations. The system reported in ref. \cite{tate} is almost the same as the system in \cite{barve2012barrier} except for two differences: First, it does not have any doping. Second, instead of AlGaAs alloys we have only GaAs material. Figure \ref{Optical_fig} shows the experimental and simulation results of the optical transition of the SK-QD system reported in ref. \cite{tate}. The optimization of the anharmonic strain model has greatly improved the simulation results. Increasing the In mole fraction increases the transition wavelength. To understand the reason behind this, one needs to see the effect of changing the mole fraction on the hydrostatic and biaxial strain and hence the band edges. Figure \ref{HBnew_L12_fig} shows the hydrostatic and biaxial strain along two lines passing through the middle of the quantum dot in the [001] and [110] directions for different cases of In mole fractions. As shown in these figures, the hydrostatic and biaxial strain change with the In mole fraction in the same way they change with diameter; increasing the In mole fraction results in an increase in the magnitude of the biaxial strain and a decrease in the magnitude of the hydrostatic strain. This leads to a lower energy gap and larger absorption wavelength, as shown in Figure \ref{Optical_fig}.

\section{Conclusion}
In this paper, a detailed theoretical study of the optical absorption and strain behavior in self-assembled quantum dots has been presented. Self-assembled quantum dots are highly strained heterostructures, and a rigorous atomistic strain model is needed to accurately calculate the electronic states in the system. We have described an optimization procedure of the anharmonic strain model, which has greatly improved the capability of simulations to reproduce experimental optical characteristics. The simulation is able to reproduce the experimental results with an error less than 1$\%$. The optimized strain model has been implemented in NEMO5 and has been used to simulate characteristics of an InAs/GaAs/AlAs quantum dot systems. The inplane-polarized light absorption is more significant than the perpendicularly polarized light absorption. 
Increasing the dot diameter results in shifting the peaks towards longer wavelengths, while
increasing the dot height doesn't seem to have a significant effect on wavelength. In case of changing the diameter, changes in band gap and confinement energies work with each others, while in case of changing the height, the changes in band gap and confinement energies work against each others.
Increasing the In mole fraction in the strain controlling layer works in the same way as increasing the dot diameter in terms of changing the strain which leads to longer absorption wavelengths. 

In conclusion, the method presented here provides a way to incorporate the inhomogeneous environment of QDs in simulations by taking into account device geometry and quantum confinement, alloy disorder, electrostatics, and spatially varying strain distribution. Such details are needed to interpret and guide experimental measurements and device design with quantitative accuracy.

\section{Acknowledgment}
This work was supported by AFRL award No. SUB1122439-001 (AMMTIAC DO 48) to the Network for Computational Nanotechnology at Purdue University. This research is also part of the Blue Waters sustained-petascale computing project, which is supported by the National Science Foundation (award number ACI 1238993) and the state of Illinois. Blue Waters is a joint effort of the University of Illinois at Urbana-Champaign and its National Center for Supercomputing Applications. This work is also part of the "Accelerating Nano-scale Transistor Innovation with NEMO5 on Blue Waters" PRAC allocation support by the National Science Foundation (award number OCI-0832623). The authors ( MSA, JWA, and BRW) are thankful for the funding support through AFOSR Lab Task 14RY07COR  (PO: Dr. G. Pomrenke).
The use of nanoHUB.org computational resources operated by the Network for Computational Nanotechnology funded by the US National Science Foundation under grant EEC-1227110, EEC-0228390, EEC-0634750, OCI-0438246, and OCI-0721680 is gratefully acknowledged. This research was supported in part by computational resources provided by Information Technology at Purdue University, West Lafayette, Indiana.

NEMO5 developments were critically supported by an NSF Peta-Apps award OCI-0749140 and by Intel Corp. Any opinions, findings, and conclusions or recommendations expressed in this material are those of the authors and do not necessarily reflect the views of the National Science Foundation.

\bibliographystyle{ieeetr}

\section*{Figures}

\begin{figure}[H]
\centering
\includegraphics[width=20pc]{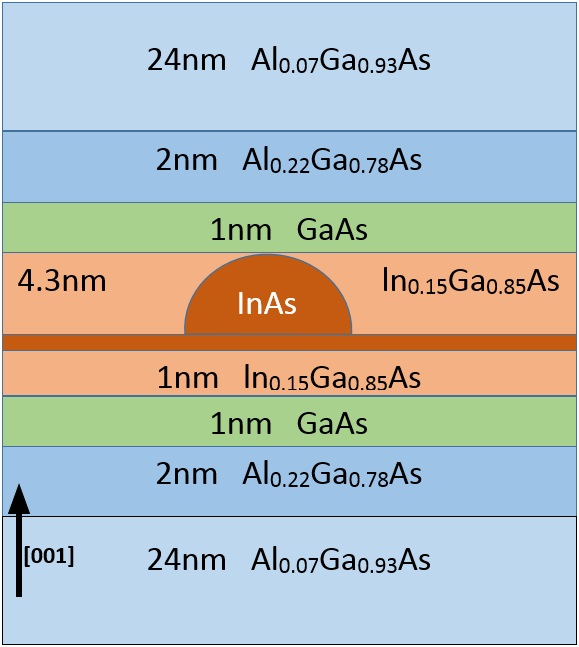}
\caption{A schematic of the measured and simulated SK-QD system. The dimension of the simulated structure is 60 nm X 60 nm X 60 nm. The quantum dot is a dome shaped InAs with base diameter of 20 nm and height of 5 nm, while the wetting layer is 2 monolayers. The strain controlling layer of In$_{0.15}$Ga$_{0.85}$As is sandwiched  between two 1 nm layers of GaAs, and two 2 nm layers of  Al$_{0.22}$Ga$_{0.78}$As. The rest of the structure is made of Al$_{0.07}$Ga$_{0.93}$As.}
\label{SKQD_fig}
\end{figure}

\begin{figure}[H]
\centering
\includegraphics[width=20pc]{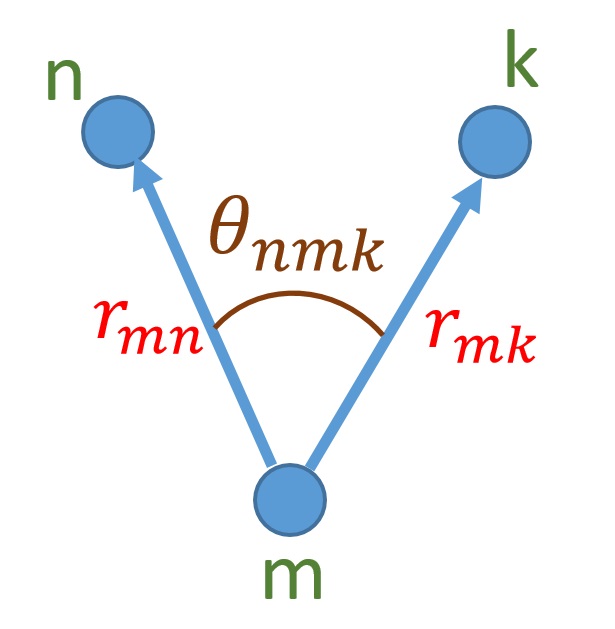}
\caption{Bond lengths and bond angle for three neighboring atoms m, n, and k.}
\label{interactions_fig}
\end{figure}

\begin{figure}[H]
\centering
\includegraphics[width=20pc]{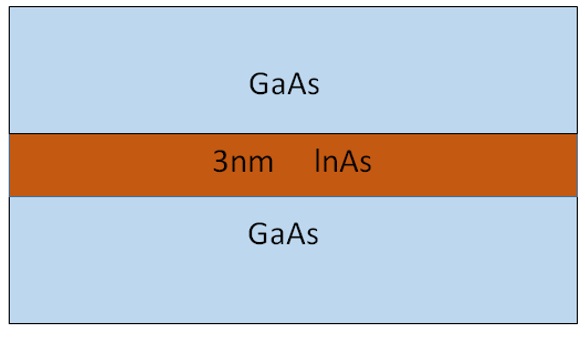}
\caption{An InAs/GaAs Quantum well of thickness 3 nm used for the optimization of the anharmonic strain model. .}
\label{QW_fig}
\end{figure}

\begin{figure}[H]
\centering
\includegraphics[width=20pc]{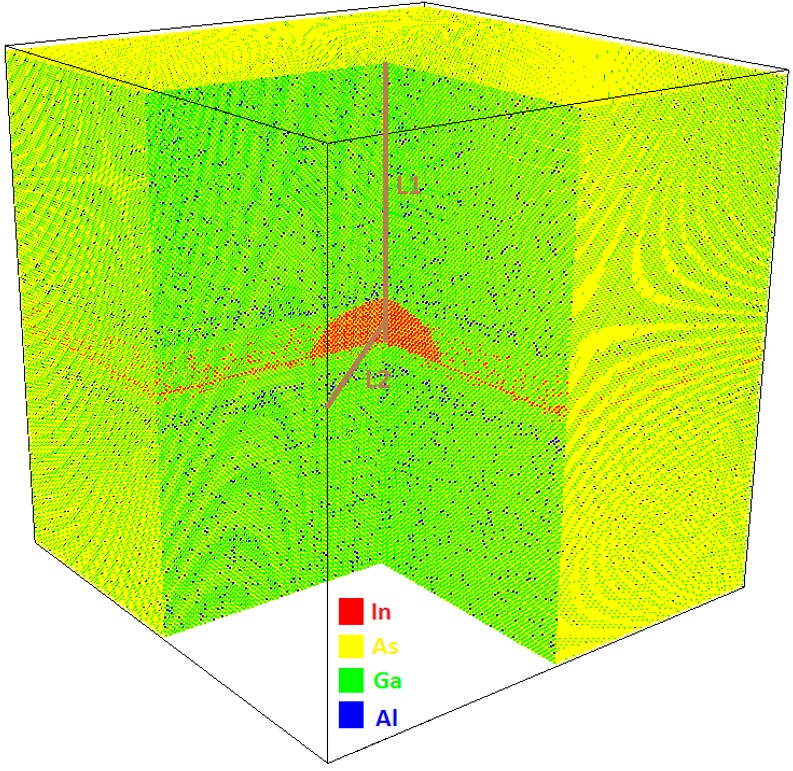}
\caption{The atomistic grid of the simulated SKQD showing the different random alloy regions. The number of atoms used for strain simulation $\approx$ 10 Million atoms, while for band structure calculations only $\approx$ 1.5 Million atoms are needed. Lines L1 and L2 are two lines passing through the middle of the quantum dot in the directions [001] and [110] respectively.}
\label{Atomistic_fig}
\end{figure}

\begin{figure}[H]
\centering
\includegraphics[width=20pc]{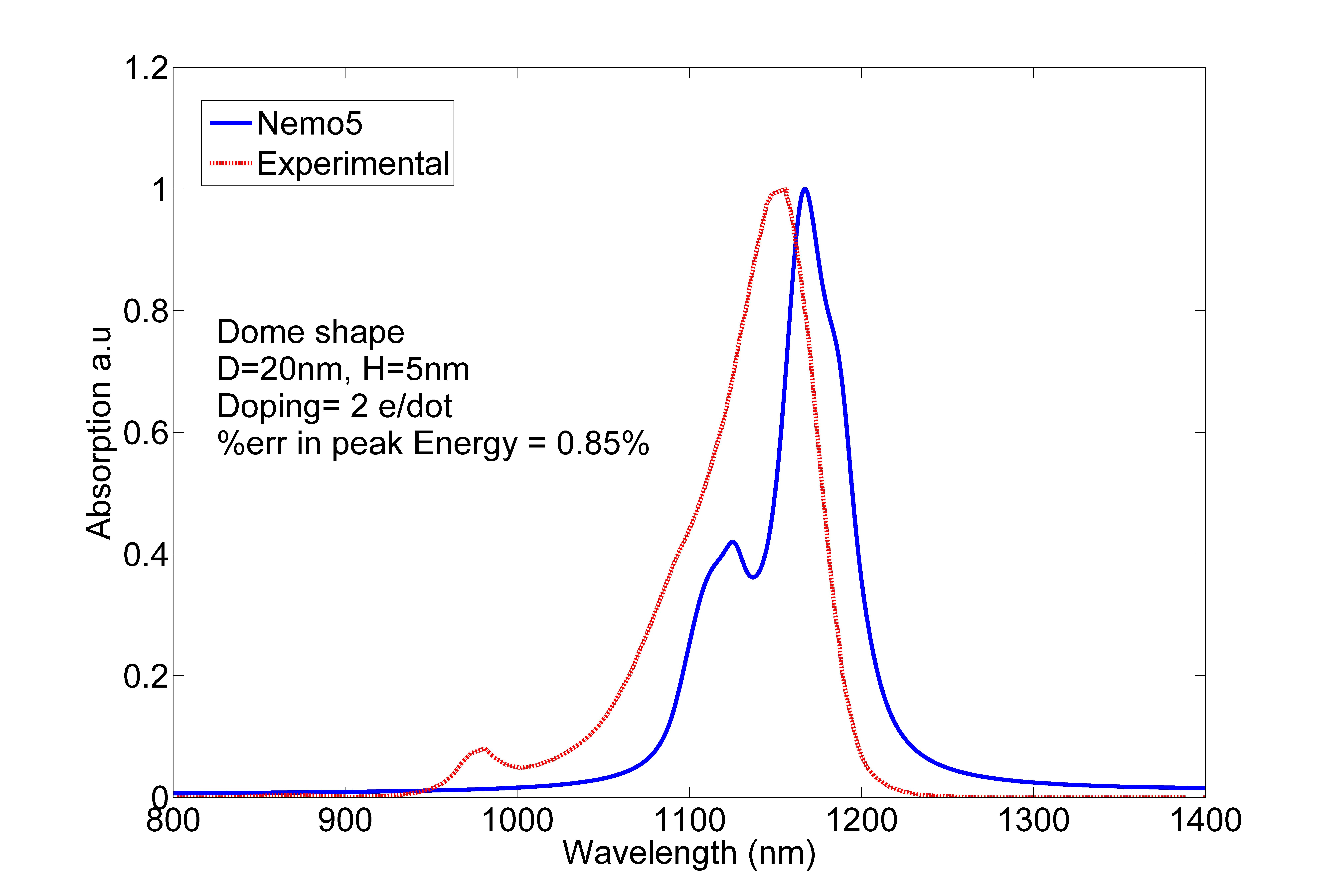}
\caption{This figure shows the calculated and measured \cite{barve2012barrier} absorption of the SK-QD system.  The  quantum dot is dome shaped with base diameter of 20 nm and height of 5 nm. The doping is 2 electrons per dot. The calculated absorption matches well with the experiment, the error is less than 1$\%$ in the absorption peak.}
\label{Experimental_fig}
\end{figure}

%

\begin{figure}[H]
\centering
\includegraphics[width=20pc]{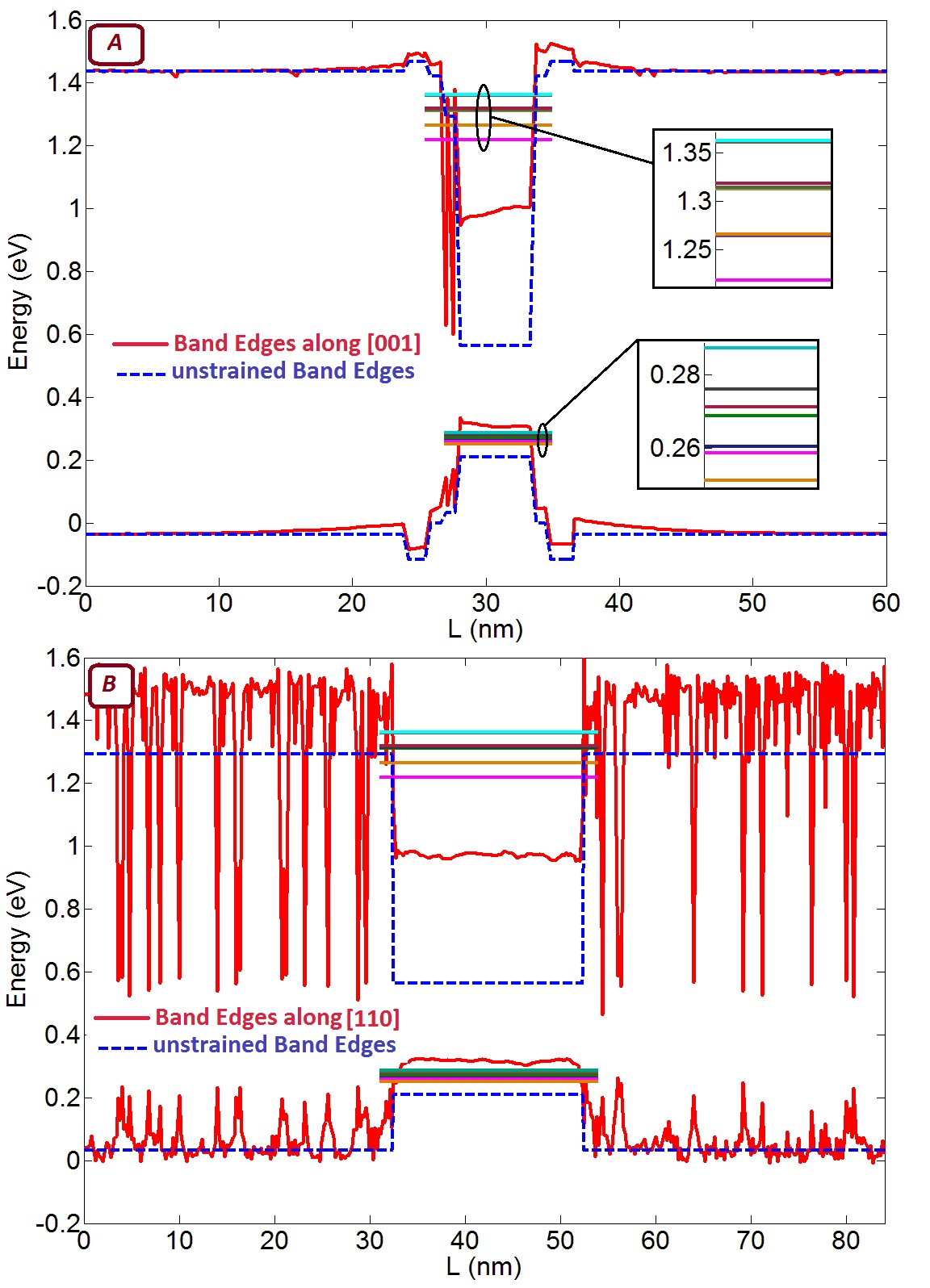}
\caption{The conduction and valence band edges (solid lines) along a line through the middle of the quantum dot in the [001] and [110] directions. The dashed lines are the band edges of the unstrained bulk materials, drawn to show the significant effect of strain on deforming the band structure. Also shown are the lowest energy bound states in the quantum dot.}
\label{LBS12_fig}
\end{figure}

\begin{figure}[H]
\centering
\includegraphics[width=20pc]{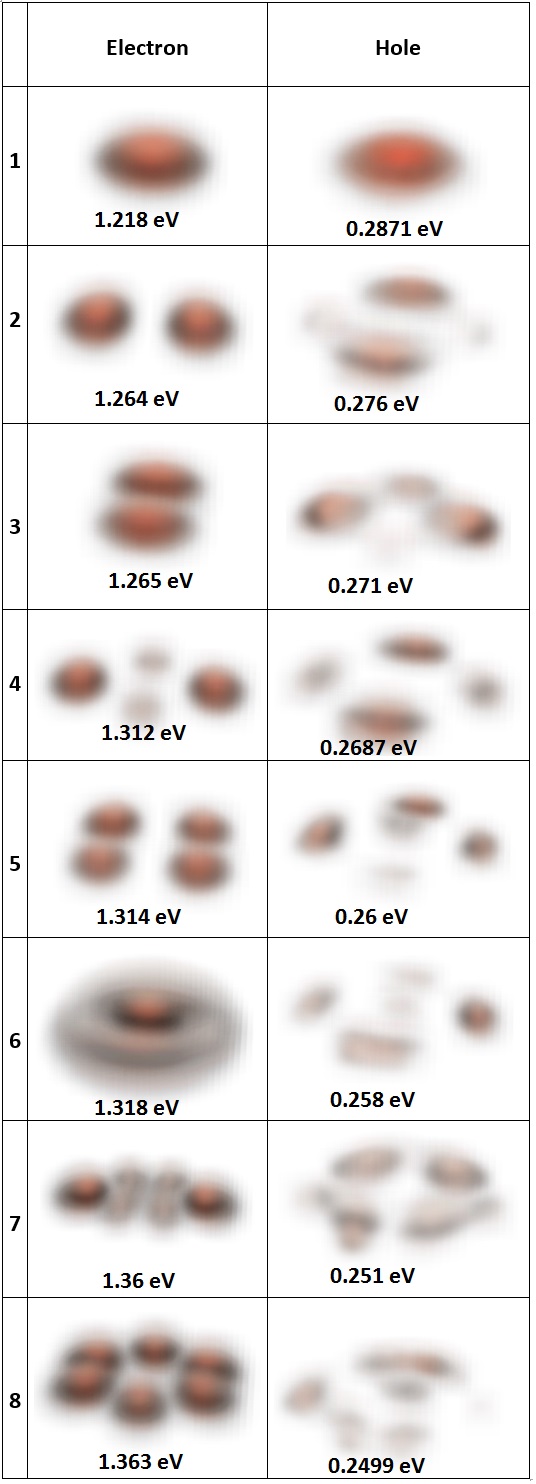}
\caption{The magnitude square of the wave functions of the electron and hole state. Plotting only the first eight electron and hole states.}
\label{States2_fig}
\end{figure}

%

\begin{figure}[H]
\centering
\includegraphics[width=20pc]{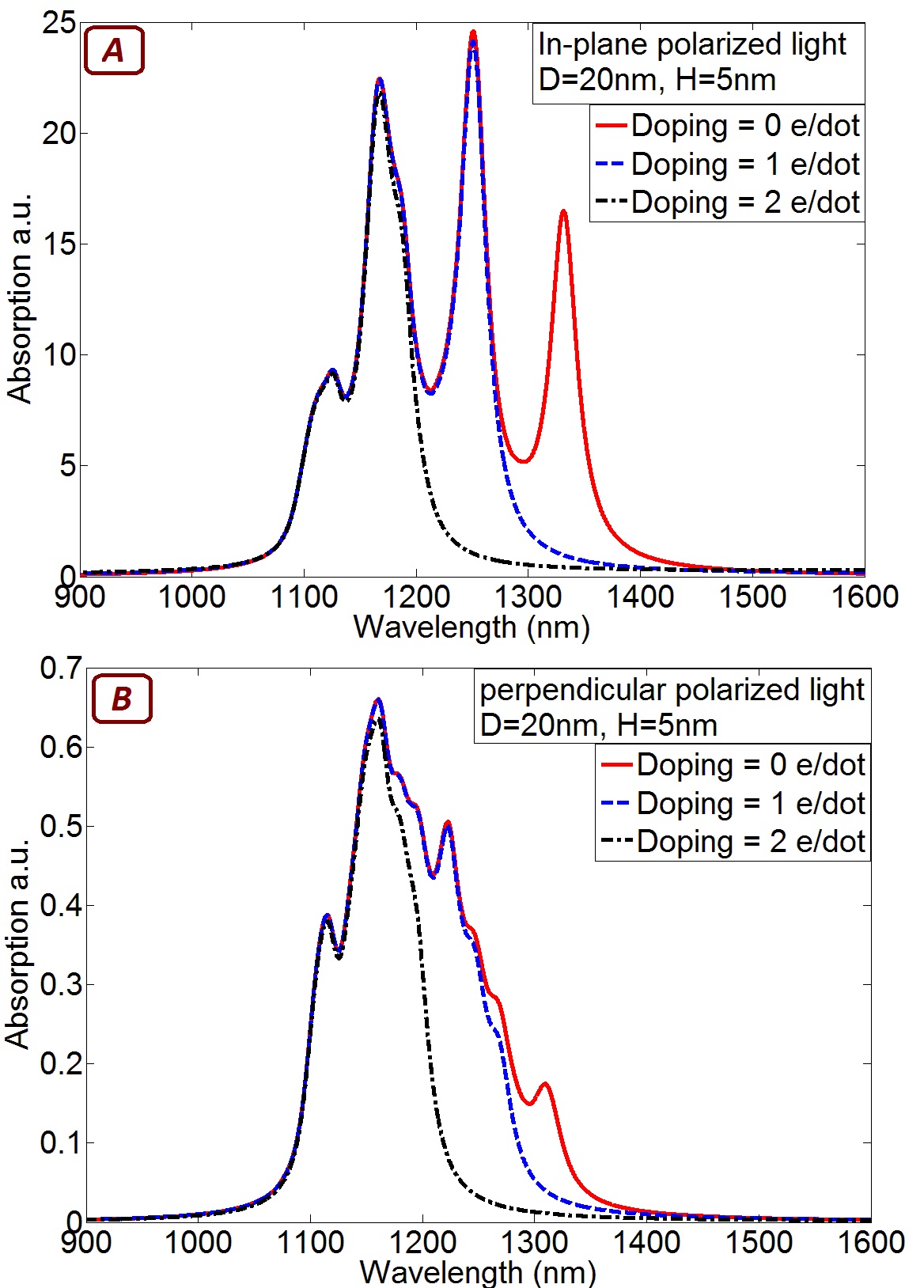}
\caption{The absorption spectrum at different cases of doping for in-plane (A) and perpendicularly (B) polarized incident light.}
\label{InPlanePerpendicular_fig}
\end{figure}

%

\begin{figure}[H]
\centering
\includegraphics[width=20pc]{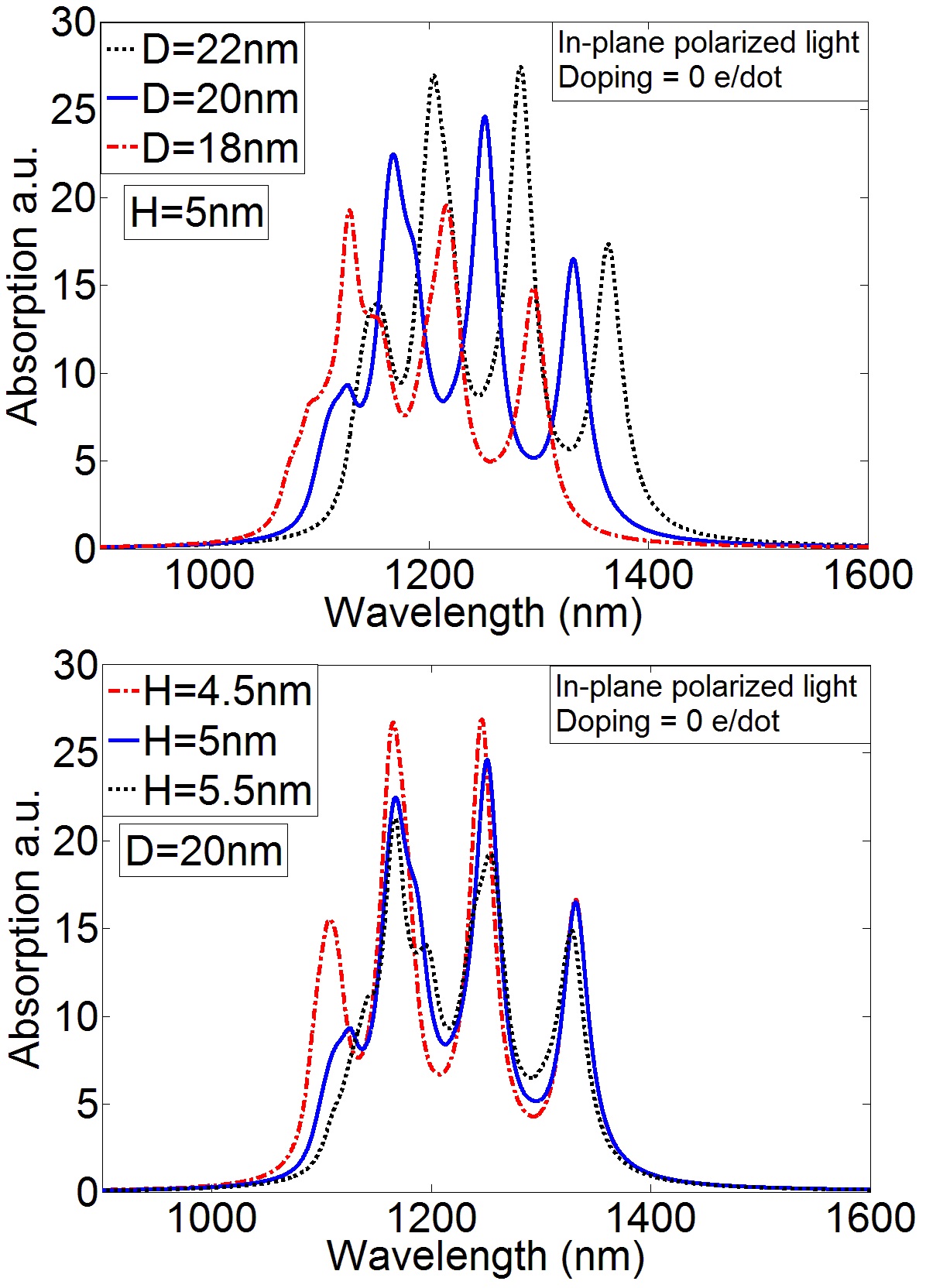}
\caption{The in-plane polarized absorption spectrum calculated for different diameters (A) and different heights (B) of the quantum dot. Increasing the dot diameter results in shifting the peaks towards longer wavelengths, while increasing the dot height doesn't seem to have a significant effect on wavelength.}
\label{DiameterHeight_fig}
\end{figure}

%
%
%

\begin{figure}[H]
\centering
\includegraphics[width=35pc]{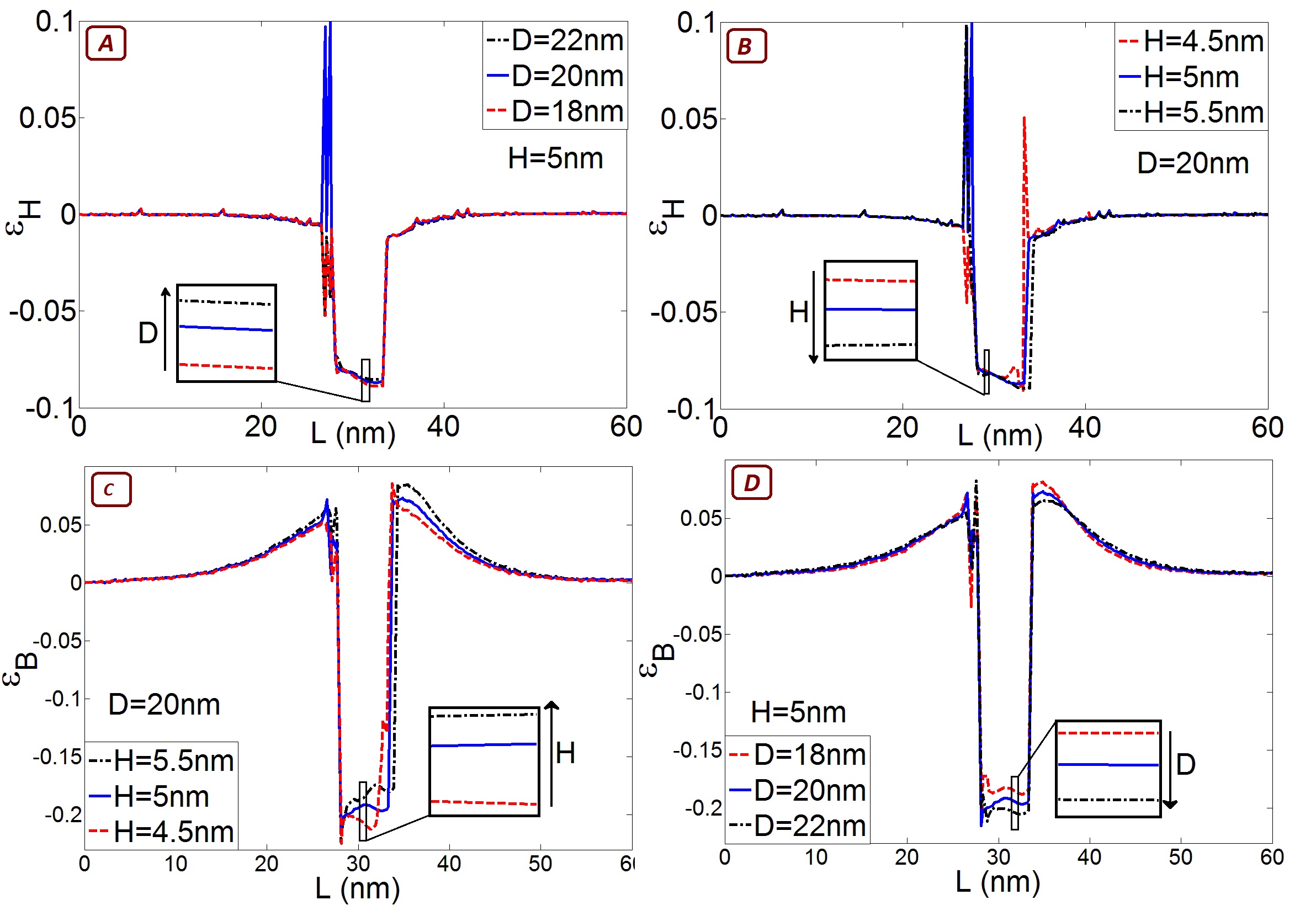}
\caption{Hydrostatic ${\varepsilon }_{H}$  and Biaxial ${\varepsilon }_{B}$ strain with different dimensions along a line through the middle of the quantum dot in the [001] direction. Figure A is ${\varepsilon }_{H}$ at different dot diameters,  Figure B is ${\varepsilon }_{H}$ at different dot heights, Figure C is ${\varepsilon }_{B}$ at different dot diameters, and  Figure D is ${\varepsilon }_{B}$ at different dot heights. The magnitude of the biaxial strain increases with increasing the diameter and decreases with increasing the height, while the hydrostatic strain behaves in the opposite way.}
\label{HB_DH_fig}
\end{figure}


\begin{figure}[H]
\centering
\includegraphics[width=20pc]{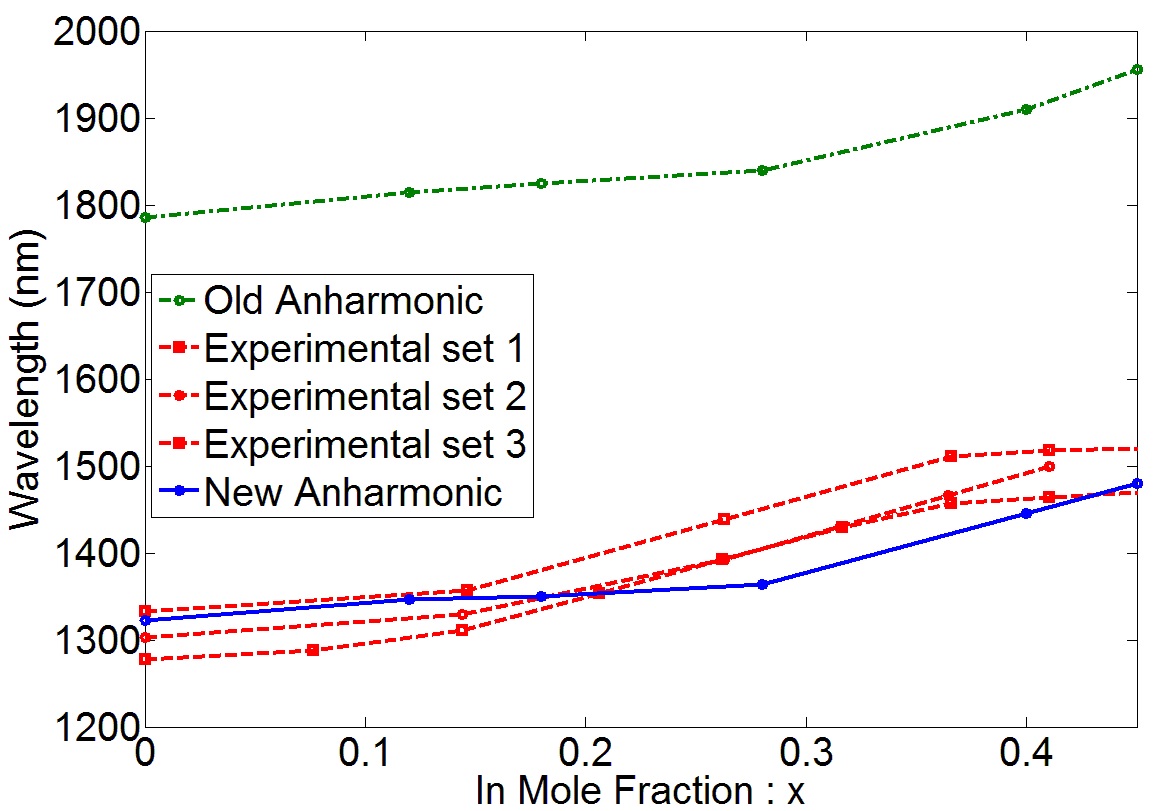}
\caption{Experimental and simulation results of the optical transition of the SK-QD system reported in ref. \cite{tate}. Increasing the In mole fraction increases the transition wavelength. The optimization of the anharmonic strain model has greatly improved the simulation results.}
\label{Optical_fig}
\end{figure}

%
%
%

\begin{figure}[H]
\centering
\includegraphics[width=40pc]{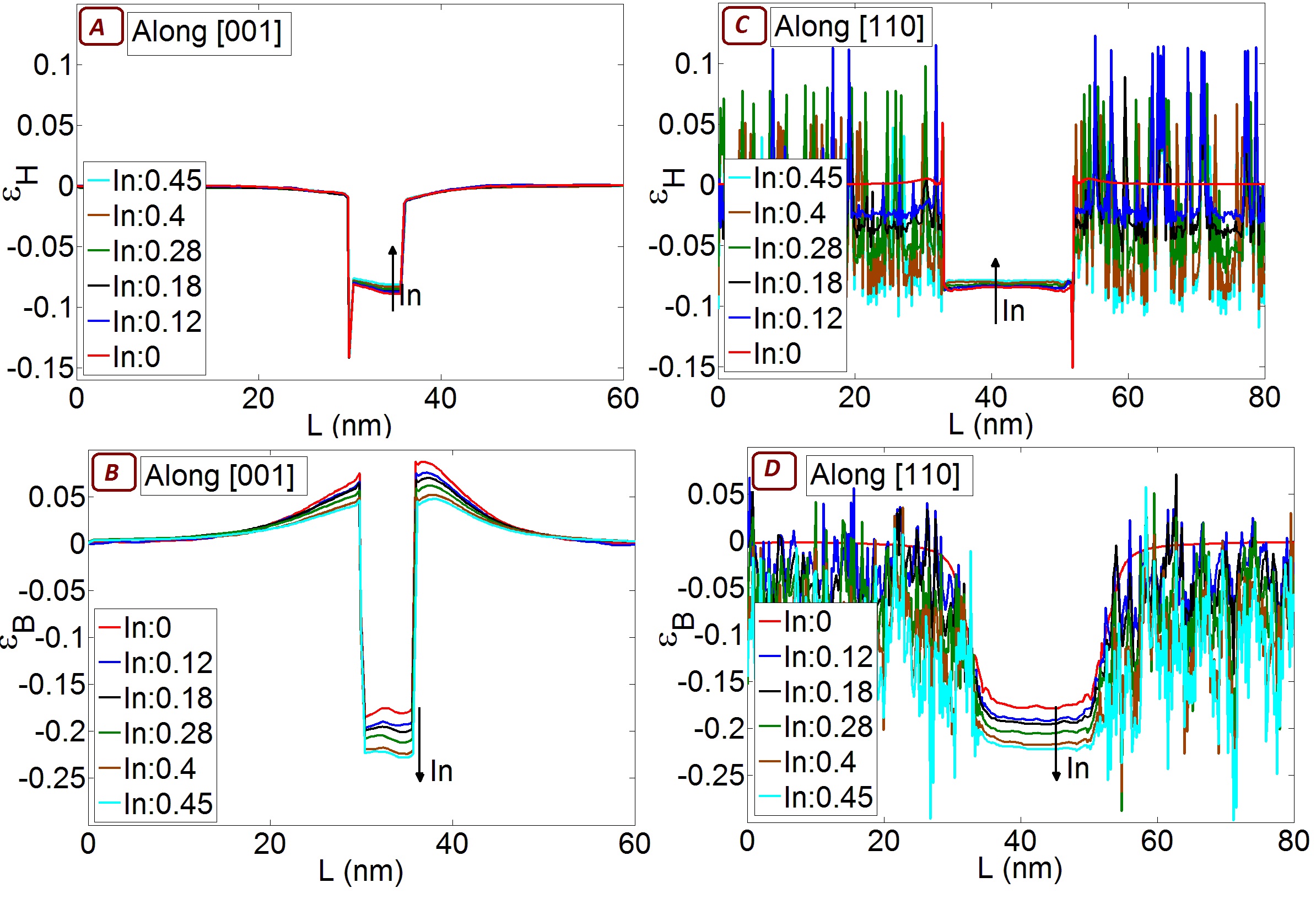}
\caption{Hydrostatic and biaxial strain with different In mole fraction along two line through the middle of the quantum dot in the [001] and [110] direction.}
\label{HBnew_L12_fig}
\end{figure}


\section*{Figure captions}
\begin{itemize}
  \item Figure \ref{SKQD_fig}: A schematic of the measured and simulated SK-QD system. The dimension of the simulated structure is 60 nm X 60 nm X 60 nm. The quantum dot is a dome shaped InAs with base diameter of 20 nm and height of 5 nm, while the wetting layer is 2 monolayers. The strain controlling layer of In$_{0.15}$Ga$_{0.85}$As is sandwiched  between two 1 nm layers of GaAs, and two 2 nm layers of  Al$_{0.22}$Ga$_{0.78}$As. The rest of the structure is made of Al$_{0.07}$Ga$_{0.93}$As.
  \item Figure \ref{interactions_fig}: Bond lengths and bond angle for three neighboring atoms m, n, and k.
  \item Figure \ref{QW_fig}: An InAs/GaAs Quantum well of thickness 3 nm used for the optimization of the anharmonic strain model.
  \item Figure \ref{Atomistic_fig}: The atomistic grid of the simulated SKQD showing the different random alloy regions. Lines L1 and L2 are two lines passing through the middle of the quantum dot in the directions [001] and [110] respectively.
  \item Figure \ref{Experimental_fig}: This figure shows the calculated and measured absorption of the SK-QD system.  The  quantum dot is dome shaped with base diameter of 20 nm and height of 5 nm. The doping is 2 electrons per dot. The calculated absorption matches well with the experiment, the error is less than 1$\%$ in the absorption peak.
  \item Figure \ref{LBS12_fig}: The conduction and valence band edges (solid lines) along a line through the middle of the quantum dot in the [001] and [110] directions. The dashed lines are the band edges of the unstrained bulk materials, drawn to show the significant effect of strain on deforming the band structure. Also shown are the lowest energy bound states in the quantum dot.
  \item Figure \ref{States2_fig}: The magnitude square of the wave functions of the electron and hole state. Plotting only the first eight electron and hole states.
  \item Figure \ref{InPlanePerpendicular_fig}:  The absorption spectrum at different cases of doping for in-plane (A) and perpendicularly (B) polarized incident light.
  \item Figure \ref{DiameterHeight_fig}: The in-plane polarized absorption spectrum calculated for different diameters (A) and different heights (B) of the quantum dot. Increasing the dot diameter results in shifting the peaks towards longer wavelengths, while increasing the dot height doesn't seem to have a significant effect on wavelength.
  \item Figure \ref{HB_DH_fig}: Hydrostatic ${\varepsilon }_{H}$  and Biaxial ${\varepsilon }_{B}$ strain with different dimensions along a line through the middle of the quantum dot in the [001] direction. Figure A is ${\varepsilon }_{H}$ at different dot diameters,  Figure B is ${\varepsilon }_{H}$ at different dot heights, Figure C is ${\varepsilon }_{B}$ at different dot diameters, and  Figure D is ${\varepsilon }_{B}$ at different dot heights. The magnitude of the biaxial strain increases with increasing the diameter and decreases with increasing the height, while the hydrostatic strain behaves in the opposite way.
  \item Figure \ref{Optical_fig}: Experimental and simulation results of the optical transition of the SK-QD system reported in ref. \cite{tate}. Increasing the In mole fraction increases the transition wavelength. The optimization of the anharmonic strain model has greatly improved the simulation results.
  \item Figure \ref{HBnew_L12_fig}: Hydrostatic and biaxial strain with different In mole fraction along two line through the middle of the quantum dot in the [001] and [110] direction.
\end{itemize}

%
%
%
%
%
%
%
%

\end{document}